\documentclass[twocolumn,floatfix,prl,aps,showpacs]{revtex4}
\usepackage{graphicx,color,amsmath,amssymb}
\usepackage{hyperref}

\newcommand{\be}{\begin{equation}}
\newcommand{\ee}{\end{equation}}

\begin{document}
\title{Possible realization of a chiral p-wave paired state in a two component system}
\author{Sutirtha Mukherjee,$^1$ J. K. Jain,$^2$ and Sudhansu S. Mandal$^1$}
\affiliation{$^1$Department of Theoretical Physics, Indian Association for the Cultivation of Science, Kolkata 700032, India}
\affiliation{$^2$Physics Department, 104 Davey Laboratory, Pennsylvania State University, University Park, PA 16802, USA}
\date{\today}
\begin{abstract} 

There is much interest in the realization of systems with p-wave pairing in one dimension or chiral p-wave pairing in two dimensions, because these are believed to support Majorana modes at the ends or inside vortices. We consider a two component system of composite fermions and provide theoretical evidence that, under appropriate conditions, the screened interaction between the minority composite fermions is such as to produce an almost exact realization of p-wave paired state described by the so-called anti-Pfaffian wave function. This state is predicted to occur at filling $\nu=3/8$ or 13/8 in GaAs when the Zeeman energy is sufficiently small, and at $\nu=\pm 3/8$ or $\pm 13/8$ in single layer graphene when either the Zeeman or the valley splitting is sufficiently small. 
\end{abstract}.

\maketitle

Much attention has recently been devoted to topological phases of matter, in particular on the realization of a one-dimensional p-wave superconductor \cite{Kitaev03} that traps Majorana modes at the ends, or a two-dimensional chiral p-wave superconductor \cite{Volovik99,Read00} whose Abrikosov vortices can support Majorana modes. Many possible experimental realizations have been suggested \cite{Theory1,Theory2,Theory3,Theory4} for this purpose, and some evidence \cite{Expt1,Expt2,Expt3} for Majorana modes has been reported at the end of one-dimensional wires. It has been proposed by Moore and Read \cite{Moore91} that the 5/2 fractional quantum Hall effect (FQHE) is a realization of a chiral p-wave superconductor of spin-polarized composite fermions, and thus provides a platform for Majorana modes obeying non-Abelian braid statistics \cite{Read00,Greiter91,Nayak96,Scarola00}. In spite of experimental efforts \cite{Nonab_expt1,Nonab_expt2,Nonab_expt3} to test this prediction, the situation has remained unclear. One possible source of complication is that the Majorana modes are best motivated for the so-called Pfaffian \cite{Moore91} or the anti-Pfaffian \cite{Levin07, Lee07, Bishara09} wave function, but the actual Coulomb 5/2 FQHE state is not accurately represented by either of these wave functions \cite{Scarola02}. It would therefore be important to ask if a better realization of the Pfaffian wave function exists. One can construct a model 3-body interaction \cite{Greiter91} that has the Pfaffian wave function as its exact solution (in this model the only interaction is that experienced by three particles in their smallest relative angular momentum state; there is no interaction in the two particle channel), but an experimental realization of such a model is difficult and has not yet been achieved. 

Composite fermions  \cite{Jain89} (CFs) are bound states of electrons and an even number ($2p$) of vortices, formed when electrons are confined to the lowest Landau level (LL). They experience an effective magnetic field and form Landau-like levels. Their filling factor $\nu^*$ is related to the electron filling factor $\nu$ by the relation $\nu=\nu^*/(2p\nu^*+1)$. We consider in this work CFs carrying two vortices each (denoted $^2$CFs) at filling factor $\nu^*=1/2$. Can they form a paired state in the same manner as electrons in the second LL do at filling factor $\nu=1/2$? The answer is negative, because the interaction between these composite fermions has a sufficiently strong short range part that they capture two more vortices and form a Fermi sea of $^4$CFs (which appears at $\nu=1/4$), just as electrons at $\nu=1/2$ in the lowest LL capture two vortices to form a $^2$CF Fermi sea \cite{Halperin93,Kalmayer92}. However, let us now consider $^2$CFs at $\nu^*=3/2$ with $\nu^*_{\uparrow}=1$ and $\nu^*_{\downarrow}=1/2$, where the subscripts label the two components, which we will generically refer to as ``spin." This state has spin polarization of $\gamma=(\nu^*_{\uparrow}-\nu^*_{\downarrow})/(\nu^*_{\uparrow}+\nu^*_{\downarrow})=1/3$. The interaction between the minority spin composite fermions is softened due to screening by the majority spin $^2$CFs. Our calculations presented below make a strong case that the screened interaction is such that it produces for the minority spin composite fermions an almost exact realization of the anti-Pfafian wave function. In GaAs, this state will manifest as a partially spin polarized incompressible FQHE state at $\nu=3/8$ or its hole partner at $\nu=13/8$. In graphene, this state can be realized at $\nu=\pm 3/8$ and $\pm 13/8$ in the limit where either the spin or the valley symmetry is approximately valid. A definitive observation of any of these FQHE sates or a measurement of its spin polarization has so far been lacking, although some evidence for a developing FQHE at 3/8 was seen by Pan {\em et al.} \cite{Pan03} and more recently by Bellani {\em et al.}\cite{Bellani10} in GaAs quantum wells.  On the theoretical front, a previous study by  Scarola {\em et al.} \cite{Scarola02} investigated this problem by modeling the composite fermions in the spin-reversed sector as interacting with an effective two-body interaction, which was obtained by a microscopic calculation \cite{Sitko96,Lee02,Wojs10}. However, the 2-body interaction model  is not fully reliable for several reasons  \cite{Mukherjee12,Mukherjee13}: the physics is likely driven by very subtle energy scales where three and higher body terms can play a significant role; the 2-body inter-CF model assumes perfect particle-hole symmetry for $^2$CFs, which is in general not valid (an explicit violation is seen below); it fails to discriminate between the Pfaffian (Pf) and the anti-Pfaffian (APf) states; it also does not give the energy of the 3/8 state, and therefore does not allow a determination of the parameter region where this state is favored. 

A study of this issue requires an accurate quantitative theory that can capture the subtle physics of pairing.  We employ the method of composite fermion diagonalization (CFD) \cite{Mandal02}, which 
has been shown \cite{Mukherjee13,Mukherjee12} to be extremely precise in this filling factor range. In this method, we work in the spherical geometry that considers $N$ electron on the surface of a sphere exposed to a radial magnetic field with total flux $2Q$ in units of the flux quantum $\phi_0=hc/e$.  We begin with a determination of the basis states $\left\{\Phi_{1/2}^{\downarrow, (L,\alpha)} \right\}$ ($\alpha$'s label distinct basis functions in a given total angular momentum sector $L$) of the degenerate ground states of $N_\downarrow$ noninteracting spin-down fermions at flux $2Q^* = 2N_\downarrow - \lambda$, which corresponds to filling 
$\nu_\downarrow^* = 1/2$; $\lambda$ is a constant ``shift" that depends on the specific model for the state.
We next composite-fermionize this basis to construct a correlated CF basis
\begin{equation}
 \left\{ \Psi_{3/8}^{\text{CF},(L,\alpha)} \right\} =  \left\{ J^2 \Phi_1^\uparrow[\{u_r,v_r\}] \Phi_{1/2}^{\downarrow,(L,\alpha)} [\{u_j,v_j\}] \right\},
 \label{eq1}
\end{equation}
with
\begin{equation}
 J^2 = \prod_{r<s}^{N_\uparrow} (u_rv_s-u_sv_r)^2 \prod_{i<j}^{N_\downarrow} (u_iv_j-v_iu_j)^2 \prod_{r,j}^{N_\uparrow,N_\downarrow} (u_rv_j-v_ru_j)^2,
\label{Jastrow}
\end{equation}
The particle indices $\{r,s\}$ ($\{i,j\}$) refer to the $N_\uparrow$ ($N_\downarrow$) number of up-spin (down-spin) composite fermions. 
The spinor coordinates are defined as $u = \cos (\theta/2) \exp (-i\phi/2)$ and $v=\sin (\theta/2) \exp (i\phi/2)$ in terms of the spherical
coordinates\cite{Haldane83} $\theta$ and $\phi$. Multiplication by the Jastrow factor $J^2$ describes the attachment of two quantum vortices to each electron. Here $N_\uparrow = 2Q^*+1$, $N = N_\uparrow + N_\downarrow$ is the total number of electrons, and the total flux
\begin{equation}
 2Q = \frac{8N - \lambda -8}{3} \, .
\end{equation}
The value $\lambda = 3\, (-1)$ refers to the Pf (APf) shift. Finally, the full Coulomb Hamiltonian is diagonalized separately in each $L$ sector within this restricted CF-basis, evaluating the Coulomb matrix elements by the Monte Carlo method. To obtain a desirable accuracy of the energies, we have had to perform an unusually large number of Monte Carlo steps (400--2000 Million for each of the systems). The spin polarization of the state is given by $\gamma = (\nu^*_\uparrow - \nu^*_\downarrow)/(\nu^*_\uparrow + \nu^*_\downarrow) = 1/3$.

As we shall see, distinguishing between different possibilities at 3/8 requires a study of fairly large systems. The lowest LL Hilbert space dimensions in various total orbital angular momentum $L$ sectors for several $(N,2Q)$ systems corresponding to Pf and APf shifts are tabulated in Table~\ref{Exact}; these are beyond the reach of computer calculations. The dimensions of CF basis $\left\{ \Psi_{3/8}^{\text{CF},(L,\alpha)} \right\}$ are exponentially small (see Table~\ref{CFD}), which allows us to treat larger values of $N$.

\begin{widetext}
\begin{table}[h!]
\caption{Exact Hilbert space dimensions for various systems $(N,2Q)$ corresponding to Pf and APf shifts in various $L$ sectors. The exact numbers are quoted for the $(N,2Q)=(14,35)$ and (16,39), and approximate numbers for larger systems.}
\label{Exact}
\setlength{\tabcolsep}{0.15em}
\centering
\footnotesize 
\vspace{2ex}
\hspace{-0.5cm}\begin{tabular}{c|| rrrrrrrrr}
\hline\hline
$(N,2Q)$&$L=0$&1&2&3&4&5&6&7&8\,\\ 
\hline
\hline

$(14,35)$&68061370 & 203942248&339434629 &474057321 & 607668097
         &739793185 &870299200 & 998724540 &1124948790 
         \\
$(16,39)$&8634541516 & 25888613188 &43101482607 & 60243279869 & 77288105284
        &94206535507&110973258761 & 127559585561 & 143941070616 
         \\
$(20,51)$& $3.23 \times 10^{13}$ & $9.70 \times 10^{13}$ & $1.61 \times 10^{14}$ & $2.26 \times 10^{14}$ & $2.90 \times 10^{14}$
         & $3.54 \times 10^{14}$ & $4.18 \times 10^{14}$ & $4.81 \times 10^{14}$ &$5.44 \times 10^{14}$ 
         
%$(20,51)$&32348581513536 &97019925956992 &161614103573888 & 226079610950880& 290365366978272
%         &354420268548768 & 418193766919520&481635423001312 &544695481865600 & 607324427439232       
       \\
$(22,55)$&$5.05 \times 10^{15}$ &$1.51 \times 10^{16} $ &$2.52 \times 10^{16}$ &$3.53 \times 10^{16}$ & $4.53 \times 10^{16}$
         &$5.54 \times 10^{16}$ &$6.54 \times 10^{16}$ &$7.53 \times 10^{16}$ &$8.52 \times 10^{16}$  
%$(22,55)$&5054280494694400 &15159712871956480 &25255765764464640 &35336194600067072 & 45394773792169984
%        &55425296655663104 &65421588035051520 &75377504156205056 &85286945160912896 & 95143854861213696       
         \\
         
\hline\hline
\end{tabular}
\end{table}
\end{widetext}

\begin{table}[h]
\caption{Dimensions of correlated CF basis for various systems $(N,2Q)$ in different $L$-sectors. 
These are equal to the ground state degeneracy for $N_\downarrow$ noninteracting particles at $2Q^*$.
Systems related by the particle-hole symmetry have the same Hilbert-space dimensions.
}
\label{CFD}
\footnotesize
\begin{tabular}{c rrrrrrrrr}
\hline\hline
$(N,2Q)$&$L=0$&1&2&3&4&5&6&7&8\,\\
\hline
\hline
$(14,35)$ \& $(16,39)$&2&0&2&1&3&1&3&1&2\,\\
$(20,51)$ \& $(22,55)$&4&1&7&5&11&7&13&9&13\,\\
$(26,67)$ \& $(28,71)$&12&10&32&30&51&48&66&61&77\,\\
\hline\hline
\end{tabular}
\end{table}

\begin{table}[h]
\caption{This table compares the Pf and the APf trial wave functions $\Psi_{3/8}^{\rm{trial-Pf/APf}}$ of Eq.~\ref{overlap} with the ground states at those $(N,2Q)$ values obtained by CF diagonalization, labeled $\Psi_{3/8}^{\rm{CFD}}$. The total number of electrons is $N$ and the corresponding fluxes are $2Q=(8N-11)/3$ for the Pf shift $(\lambda =3)$ and $2Q=(8N-7)/3$ for the APf shift $(\lambda =-1)$. The corresponding number of down-spin $^2$CFs and effective flux $2Q^*$ are also listed.
The overlap between the CFD ground state $\Psi_{3/8}^{\rm{CFD}}$ and the trial wave function $\Psi_{3/8}^{\rm{trial-Pf/APf}}$ is denoted $\langle \Psi_{3/8}^{\rm{CFD}}|\Psi_{3/8}^{\rm{trial-Pf/APf}}\rangle$. For $(N,2Q)=(28,71)$ the CFD ground state occurs at $L = 4$; the asterisk is to remind us that the overlap shown in this case is calculated with the lowest energy state in the $L=0$ sector. 
}
\label{overlappf}
\begin{tabular}{ c  c  c  c  c  c } \hline\hline
~$N$~&~$2Q$~&$N_\downarrow$~&~$2Q^*$~&~$\langle\Psi_{3/8}^{\rm{CFD}}|\Psi_{3/8}^{\rm{trial-Pf}}\rangle$ &~$\langle\Psi_{3/8}^
{\rm{CFD}}|\Psi_{3/8}^{\rm{trial-APf}}\rangle$ \\ \hline\hline
14 & 35 & 4 & 9&  -- & 0.9996(1)  \\
16 & 39 &    6 & 9 &       0.9815(1) & --  \\
 20 & 51 &   6 & 13 &          -- & 0.9999(1)  \\
22 & 55 &    8 & 13 &         0.7739(3) & -- \\
26 & 67 &    8 & 17 &        -- & 0.9969(1) \\
$28^*$& 71&  10 & 17 &      0.9612(3) & --\\ \hline\hline
\end{tabular}
\end{table}

\begin{figure}[h]
\centering
\includegraphics[width=8.5cm,angle=0]{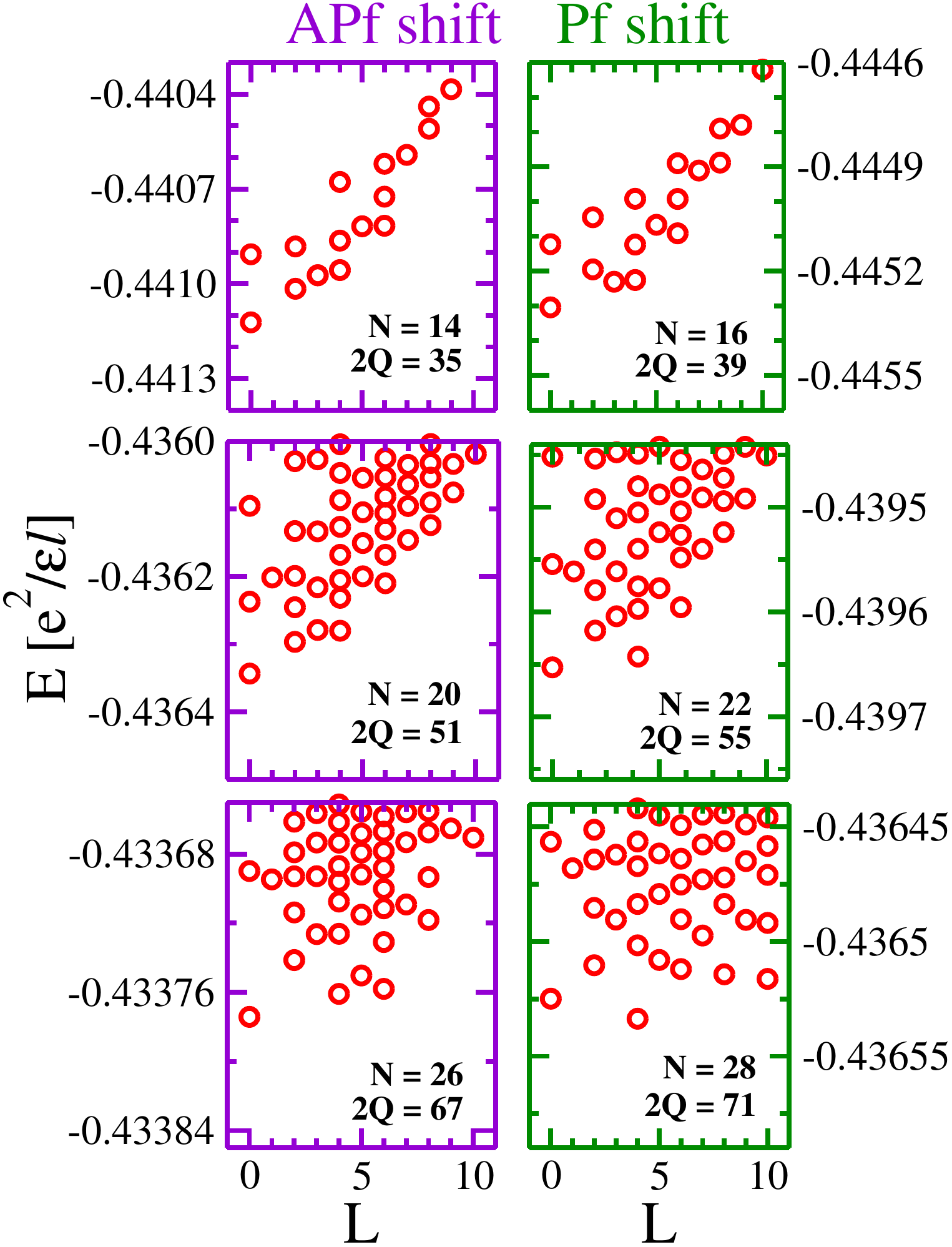}
\caption{(Color online) The low energy spectra obtained by CF diagonalization at both ``Pf shift" (right panels) and ``APf shift" (left panels) at partially spin polarized $\nu=3/8$, for systems with various values of $N$ and $2Q$ shown on the panels.  The statistical error, estimated from Metropolis Monte Carlo calculation is less than the diameter of the circle. The energy per electron, $E$, includes the interaction with the background; it is quoted in the units of $e^2/\epsilon l$ where $l = \sqrt{\hbar c/eB_\perp}$ is the magnetic length, $\epsilon$ is dielectric constant of the host material, and $B_\perp$ is the component of the magnetic field along perpendicular to the plane of the system. 
}
\label{spectra}
\end{figure}

The low-lying spectra obtained by CFD for several systems are shown in Fig.~\ref{spectra} at the APf and Pf shifts (left and right columns, respectively).  The horizontal neighbors correspond to the same $Q^*$, and would have identical spectra if composite fermions satisfied particle-hole symmetry (as would be the case in the model of Ref.~ \cite{Scarola02}). An incompressible state is identified by an $L=0$ ground state which is separated from other states by a nonzero gap in the thermodynamic limit. Several important points can be noted from the spectra:  (i) A nondegenerate ground state is formed at L=0 for each of the systems considered for the APf shift, but for the Pf shift the ground state of $(N,2Q) = (28,71)$ occurs at $L=4$. Figure~1 thus disfavors a Pf type physics for partially polarized FQHE state at 3/8. (ii) A lack of particle-hole symmetry for composite fermions can be seen, as the spectra on the right panels  are
different from their corresponding spectra on the left. (iii) The accuracy and the efficiency of the CFD method is crucial for an investigation of this incompressible state. The minimum energy gap for neutral excitation is small, of the order of $\sim 0.0004\, e^2/\epsilon l$, for the systems studied here.  However, our systems are still too small for a reliable extrapolation of the excitation gap to the thermodynamic limit.

To further corroborate the APf physics, we consider the explicit trial wave functions 
\begin{equation}
 \Psi_{3/8}^{\text{trial-Pf/APf}} = J^2 \Phi_1^\uparrow [\{u_r,v_r\}] \Phi_{1/2}^{\downarrow,\text{Pf/APf}} [\{u_j,v_j\}] \, 
\label{overlap}
\end{equation}
Here $\Phi_1^\uparrow$ is the wave function of one filled LL (with spin-up particles).  The factor $\Phi_{1/2}^{\downarrow,\text{Pf/APf}}$ represents the Pf or the APf wave function at $\nu_\downarrow^*=1/2$. The $\Phi_{1/2}^{\downarrow,\text{Pf}}$ is produced by diagonalizing the 3-body interaction Hamiltonian \cite{Greiter91} $V_3 = \sum_{i<j<k}P^{(3)}_{ijk}(3Q^*-3)$ with $2Q^*=2N_\downarrow-3$, 
where $P_{ijk}^{(3)}(L)$ projects out a cluster of three particles $(i,j,k)$ with total orbital angular momentum $L$. The APf wave function
$\Phi_{1/2}^{\downarrow,\text{APf}}$ at $\nu_\downarrow^* =1/2$ with $2Q^* = 2N_\downarrow +1$ is obtained by diagonalizing particle-hole 
conjugated Hamiltonian of $V_3$ with $2Q^*=2N_\downarrow-3$.  Overlaps of the Pf and APf-type trial wave functions  $\Psi_{3/8}^{\text{trial-Pf/APf}}$ in Eq.~(\ref{overlap}) with the corresponding CFD ground states in the $L=0$ sector are calculated using Monte Carlo method and are tabulated in Table \ref{overlappf}. Most strikingly, the APf wave function has almost perfect overlaps ($>99.6\%$) with the state obtained by CFD. We believe that, taken altogether, these results make a compelling case that a partially spin polarized incompressible state at 3/8 is possible, and, if observed, it will be an excellent realization of the APf paired state.

\begin{figure}[h]
\centering
\includegraphics[width=4.6cm,angle=0]{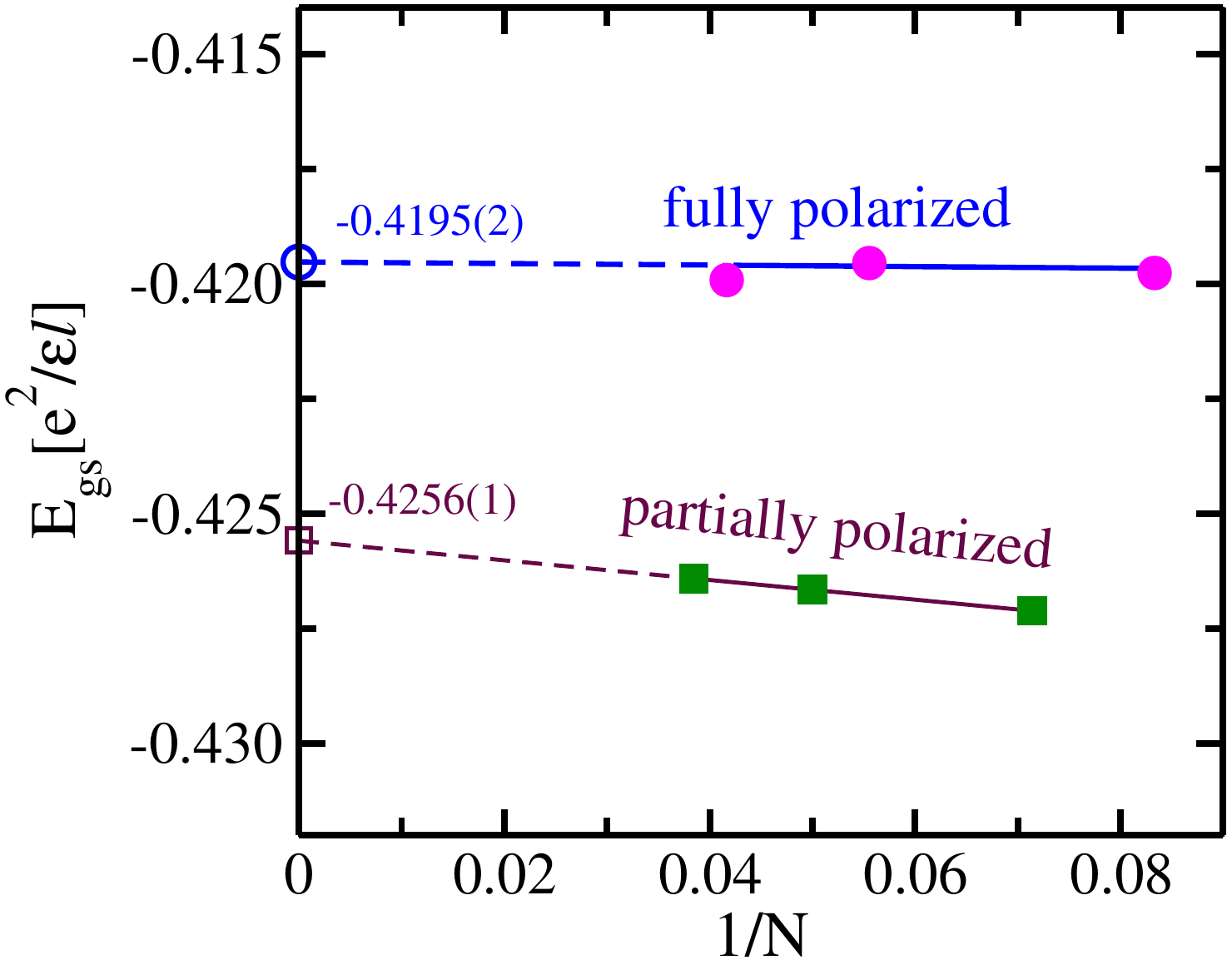}
\includegraphics[width=3.8cm]{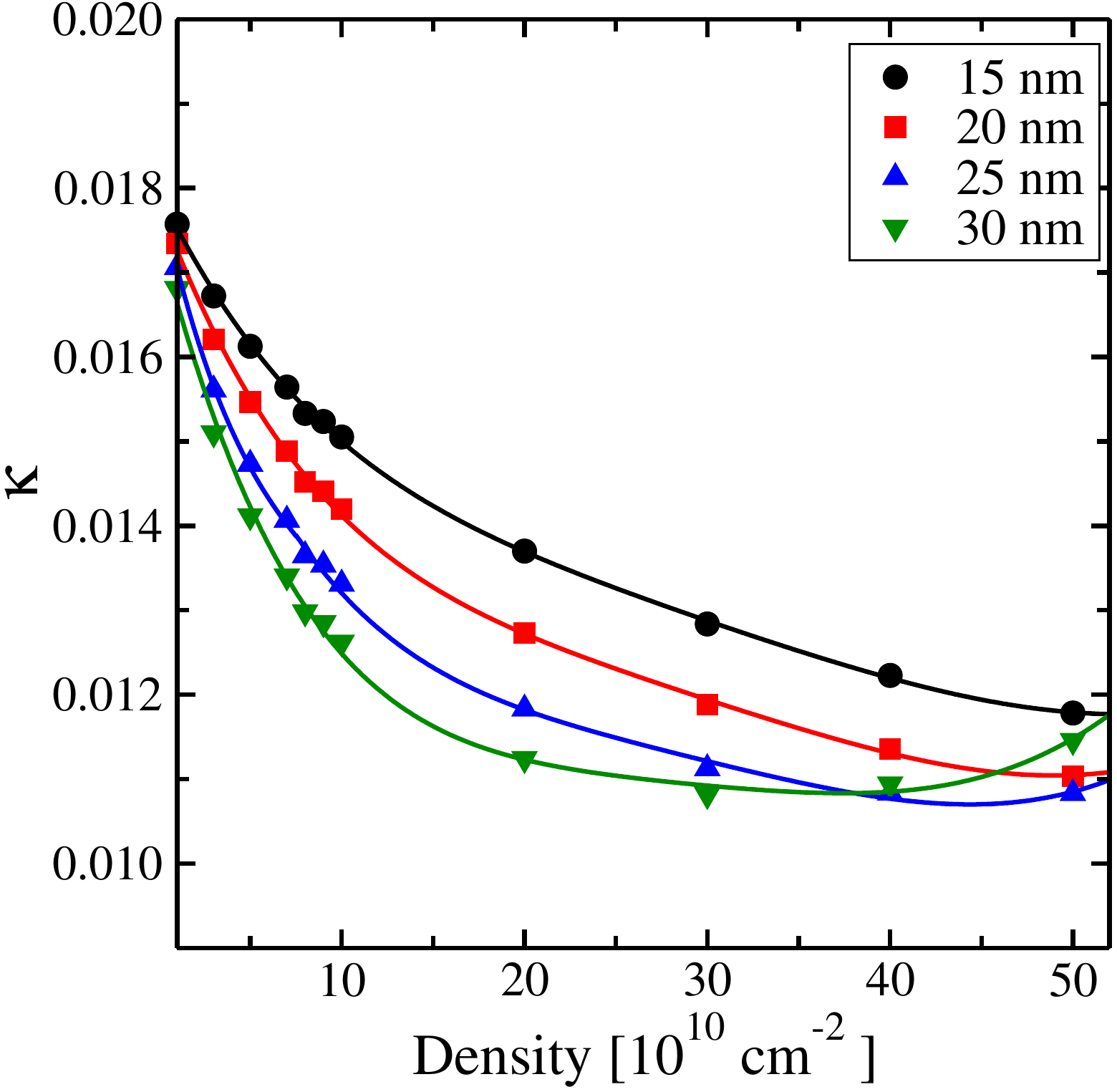}
\caption{(Color online) Left panel: The ground state energy per particle, $E_{\rm gs}$, for fully and partially polarized 3/8 states at the APf shifts (circles and squares, respectively).  A linear extrapolation in $1/N$, where $N$ is the number of particles, produces the ground state energies (open symbols) in the thermodynamic limit.   Right panel: Dependence of the critical value of $\kappa \equiv E_{\rm Z}/(e^2/\epsilon l)$, at which a transition from the partially polarized to fully polarized state takes place (see text), on electron density for several quantum well widths indicated on the figure. Here $E_{\rm Z}$ is the Zeeman splitting, $\epsilon$ is the dielectric function of the host material, and $l$ is the magnetic length.
}
\label{gs_ener}
\end{figure}

While we have made a case that the partially spin polarized 3/8 state is the lowest energy state in the chosen spin sector, its realization requires that it be the global ground state in some parameter range. As shown in Ref.~ \cite{Mukherjee12}, an APf-type state is also possible for fully spin polarized system at $\nu=3/8$. Figure \ref{gs_ener} shows the interaction energies of the ground states (including the interaction with the background) per particle, obtained by CFD at different allowed $N$ values for the both partially and fully polarized states at the APf shift. The data for fully polarized 3/8 state are taken from Ref. \cite{Mukherjee12} (although we have improved the accuracy of the energies compared to that in Ref.~ \onlinecite{Mukherjee12}). The ground state energies of the partially and fully polarized 3/8 states in the thermodynamic limit ($N \to \infty$) are obtained respectively to be $-0.4256(1)\, e^2/\epsilon l$ and $-0.4195(2)\, e^2/\epsilon l$. (Note: To correct for a slight variation in the density with $N$ in the spherical geometry, we multiply the finite system energies by the factor $(2Q\nu/N)^{1/2}$ prior to extrapolation.) The important point is that the partially spin polarized 3/8 has lower Coulomb energy than the fully spin polarized state, and therefore is favored at sufficiently low Zeeman energy $E_{\rm Z}$ (the energy required to flip the spin of an electron). Equating the interaction energy difference per particle between the two different states with $E_{\rm Z}/3$, we predict that a phase transition from partially to fully polarized state occurs at $\kappa \equiv E_{\rm Z}/(e^2/\epsilon l) \approx 0.018$, which corresponds to a magnetic field of $\sim$10 T for parameters of GaAs and also for graphene (assuming that spin is the relevant degree of freedom for the latter). We have also estimated the quantum well width dependence of the critical value of $\kappa$ for GaAs, shown in Fig.~\ref{gs_ener}, where we have evaluated the transverse wave function in a local density approximation, which modifies the form of the interaction between electrons. The finite width corrections should be negligible in graphene. We note that extensive investigation of the spin or valley physics of the FQHE states of the form $\nu = n/(2n\pm 1)$ has been performed experimentally in GaAs and AlAs quantum wells as well as graphene \cite{Eisenstein,Duspin,Cho,Yeh,Eom,Betthausen,Kukushkin,Yusa01,Gros07,Hayakawa,Melinte,Smet01,Kraus,Smet,Tiemann,Yacoby2,Shayegan1,Shayegan2,Shayegan3,Shayegan4} and these results have been analyzed quantitatively by the CF theory \cite{Wu93,Park98,Archer13}. One may wonder about the role of CF skyrmions; these are estimated to be relevant only for very small values of $\kappa<0.007$ close to $\nu=1/3$ \cite{Kamilla96}, but are not relevant for the physics of the ground state at 3/8 \cite{Archer13}.

Many proposals \cite{Levin07,Lee07,Wen93} have been made for distinguishing between the Pf and the APf states at $\nu=5/2$.  These include: counter-propagating neutral modes at the edge of the APf \cite{Lee07};  various tunnel exponents \cite{Wen93}; thermal conductivity \cite{Levin07}; {\em etc}. These ideas were extended to fully polarized 3/8 state \cite{Mukherjee12}, and carry over to the partially polarized 3/8 state without change. To summarize, the partially spin polarized APf state at 3/8 is predicted to have charge $1/16$ excitations (an abelian vortex in the spin reversed wave function $\Phi_{1/2}^{\downarrow,{\rm Pf/APf}}$ in Eq.~\ref{overlap} has a charge equal to 1/8$^{\rm th}$ of an electron charge, as can be ascertained from standard methods; however, because of pairing physics \cite{Moore91}, the vortex can be split into two quasiholes of charge 1/16 each); Majorana modes; non-Abelian braid statistics for excitations; counter propagating edge modes (none for Pf); and thermal Hall conductivity of 1/2 (5/2 for Pf) in units of $(\pi^2k_{_B}^2 /3h)T$. 

There is an interesting difference in the edge structure of the Pf and APf 3/8 states for the partially spin polarized case. We recall that for the fully polarized Pf and APf-$3/2$ state have edge structures 3/2(Pf)-1-0 and 3/2(APf)-2-1-0 (which denote the filling factor as we go across the edge from inside out), which produce at $\nu=3/8$ the edge structures 3/8(Pf)-1/3-0 and 3/8(APf)-2/5-1/3-0, respectively. In contrast, the partially polarized Pf and APf-3/2 states will have edge structures of 3/2(Pf)-1-0 and 3/2(APf)-2(singlet)-0, where the last one follows because the filling of both up and down spins will simultaneously vanish at the edge. Upon composite fermionization, we thus find the edge structures for partially polarized Pf and APf states at 3/8 to be 3/8(Pf)-1/3-0 and 3/8(APf)-2/5(singlet)-0. The partially spin polarized APf 3/8 state is thus accompanied by a sliver of spin singlet 2/5 at the boundary. We note that the calculated critical values of $\kappa$ are very different for a spin transition in the bulk states at 2/5 ($\kappa\approx 0.011$\cite{Park98}) and 3/8 ($\kappa\approx 0.018$); our calculation suggests that there is a region of $\kappa$ where the partially polarized APf 3/8 FQHE in the bulk induces a spin singlet 2/5 at the boundary even though a spin singlet 2/5 is not stable in a bulk phase in this parameter region.  It is important to remember that in our discussion of the edge physics, we have uncritically assumed the validity of the effective theory and disregarded the possibility of edge reconstruction.

In summary, we predict that the screened interaction between the minority $^2$CFs at total effective filling $\nu^*=3/2$ will produce an excellent realization of the anti-Pfaffian wave function, representing a chiral p-wave pairing of $^4$CFs. This state is relevant for fillings 3/8 and 13/8 in GaAs and $\pm$3/8 and $\pm$13/8 in graphene. We have estimated the Zeeman energy range where it should occur. 

We thank A. W\'ojs for providing us with the Fock-space representation of Pfaffian and anti-Pfaffian wave functions, and Y.-H. Wu for useful discussions. We acknowledge financial support from CSIR, Government of India (S. M.) and the DOE grant No. DE-SC0005042 (J. K. J.).

\end{document}